\documentclass[a4paper,12pt]{article}
\usepackage{amsfonts,amsthm,amsmath,amssymb,graphicx,hyperref,youngtab}

\newcommand{\be}{\begin{equation}}
\newcommand{\bea}{\begin{eqnarray}}
\newcommand{\ee}{\end{equation}}
\newcommand{\eea}{\end{eqnarray}}

\begin{document}

\makeatletter
\@addtoreset{equation}{section}
\makeatother
\renewcommand{\theequation}{\thesection.\arabic{equation}}

\rightline{QGaSLAB-13-01,\,\,WITS-CTP-109}
\vspace{1.8truecm}

\vspace{15pt}

%%%%%%%%%%%%%%%%%

{\LARGE{  
\centerline{   \bf Large $N$ anomalous dimensions for large operators} 
\centerline {\bf  in Leigh-Strassler deformed SYM} 
}}  

\vskip.5cm 

\thispagestyle{empty} \centerline{
    {\large \bf Robert de Mello Koch$^{a}$\footnote{ {\tt robert@neo.phys.wits.ac.za}}, Jeff Murugan$^{b}$\footnote{\tt jeff@nassp.uct.ac.za}}
   {\large \bf and Nkululeko Nokwara$^{a}$\footnote{\tt Nkululeko.Nokwara@students.wits.ac.za}} }

\vspace{.4cm}
\centerline{{\it $^{a}$National Institute for Theoretical Physics ,}}
\centerline{{\it Department of Physics and Centre for Theoretical Physics }}
\centerline{{\it University of Witwatersrand, Wits, 2050, } }
\centerline{{\it South Africa } }
\vspace{.4cm}
\centerline{{\it $^{b}$The Laboratory for Quantum Gravity \& Strings}}
\centerline{{\it Department of Mathematics and Applied Mathematics,}}
\centerline{{\it University of Cape Town,}}
\centerline{{\it Private Bag, Rondebosch, 7700, South Africa}}

\vspace{1.4truecm}

%%%%%%%%%%%%%%%%%
\thispagestyle{empty}

\centerline{\bf ABSTRACT}

\vskip.4cm 

We study the large $N$ anomalous dimensions of operators in a Leigh-Strassler deformation of ${\cal N}=4$ super Yang-Mills theory.
The operators that we study have a bare dimension of order $N$ (so that the large $N$ limit is not captured by planar diagrams) and 
are AdS/CFT dual to giant gravitons.
The diagonalization of the dilatation operator factorizes into two problems.
One of these problems is solved using a double coset ansatz.
The second problem is equivalent to a set of decoupled harmonic oscillators.

\setcounter{page}{0}
\setcounter{tocdepth}{2}

\newpage

\tableofcontents

\setcounter{footnote}{0}

\linespread{1.1}
\parskip 4pt

{}~
{}~

\section{Introduction}

An interesting quantity to compute for any conformal field theory is its spectrum of anomalous dimensions. 
The computation of this spectrum entails the diagonalization of the dilatation operator.
By identifying the planar dilatation operator of ${\cal N}=4$ super Yang-Mills theory as the 
Hamiltonian of a spin system,
a rich integrable structure underlying the planar limit has been discovered\cite{mz,bks}, 
allowing tremendous progress in exploring the the AdS/CFT correspondence\cite{malda,Gubser:1998bc,Witten:1998qj}.
There is even reason to hope that the exact spectrum of anomalous dimensions can be found in the planar limit. 
See \cite{intreview} for a comprehensive recent review.
Given these developments, a somewhat natural next step is to ask if integrability is present in other large $N$ limits of
${\cal N}=4$ super Yang-Mills theory. 
Recent work suggests that this is indeed the 
case\cite{Koch:2010gp,DeComarmond:2010ie,Carlson:2011hy,gs,Koch:2011hb,deMelloKoch:2011vn,mn,DCI,pnm}.

The study of integrability in large $N$ but not planar limits of ${\cal N}=4$ super Yang-Mills theory has 
focused on operators that are AdS/CFT dual to giant gravitons\cite{mst,myers,hash}.
For these operators, as a consequence of the fact that the large $N$ and planar limits do not coincide\cite{Balasubramanian:2001nh}, computing 
correlators involves more than just summing the planar diagrams.
One way to tackle this task, is by employing representation theory of the symmetric and unitary groups
as well as the relations between them. 
In this way it has been possible to construct bases of operators that diagonalize the free field two point function to all orders in 
$1/N$\cite{cjr,Balasubramanian:2004nb,dssi,Kimura:2007wy,BHR1,BHR2,countconst,Bhattacharyya:2008rb,Kimura:2008ac,Kimura:2010tx,pasram,Kimura:2012hp}.
In this work we will focus on the basis provided by the restricted Schur polynomials\cite{Balasubramanian:2004nb,dssi,Bhattacharyya:2008rb}.
These operators mix only weakly at one loop\cite{dssii,bds} and the analytic diagonalization of the one loop dilatation operator has now been 
achieved\cite{Carlson:2011hy,gs,Koch:2011hb,DCI,pnm}, for the case that we build the operators
with a large number of adjoint scalars ($Z$ say) doped with a much smaller number of impurities, which may be another species
of scalar, fermions or covariant derivatives of the original scalar $Z$.
In this case the problem of diagonalizing the one loop dilatation operator factorizes into two diagonalization problems: one associated
with the $Z$ fields and one associated with the impurities.
The diagonalization problem associated with the impurities is solved using a double coset ansatz\cite{DCI} (see also 
\cite{Koch:2011hb,mn,pnm}).
A remarkable feature of the result of this diagonalization is that one sees very concretely the emergence of the Gauss Law
expected from the AdS/CFT dual system of a giant graviton with open string 
excitations\cite{Balasubramanian:2004nb,Berenstein:2003ah,Sadri:2003mx}. 
The diaganolization associated with the $Z$ fields has been solved in \cite{gs}. 
It reduces to the motion of particles along the real line interacting via quadratic pair-wise interaction potentials.
In this way the spectrum of the dilatation operator reduces to the spectrum of a set of decoupled oscillators, which indicates
that the system is integrable.
We expect that the simple solution of the diagonalization associated with the $Z$ fields and the
associated oscillators, relies on integrability.
In contrast to this, the Gauss Law is expected to hold for any gauge theory and consequently we expect that
the double coset ansatz will continue to be useful even when integrability is not present and also when one goes to higher loops. 
This has been verified at two loops in \cite{deMelloKoch:2012sv}.
The motivation for this project is to further explore this expectation.

It is possible to add a class of deformations, first catalogued by Leigh and Strassler\cite{ls},
that break the superconformal symmetry down to ${\cal N}=1$. 
Using different choices of the deformation, it is possible to either preserve or destroy the integrability.
These deformations thus provide the ideal laboratory for us to consider the generality of the double coset 
ansatz. After deformation, the superpotential depends on three parameters
\bea
   W=i\kappa\left[ {\rm Tr}(XYZ-qXZY)+{h\over 3}{\rm Tr}(X^3+Y^3+Z^3)\right]
\eea
In this article we will consider the simplest case of a $\beta$-deformation for which $q=e^{-2i\pi\gamma}$, $h=0$ and $\gamma$ is real.
This deformation preserves integrability \cite{Roiban:2003dw,Berenstein:2004ys,Beisert:2005if,Frolov:2005ty}. 
In the next section we will evaluate the action of the one loop
dilatation operator of the deformed theory on restricted Schur polynomials. With this result in hand, we are
able to apply the double coset ansatz and show that it continues to provide the diagonalization for the
impurity labels. The diagonalization problem for the $Z$ labels is quite different to the problem studied in \cite{gs}.
In section 3 we take a continuum limit and show in section 4 that in this limit the spectrum of the dilatation operator again reduces
to a set of decoupled oscillators. Section 5 is reserved for discussion of our results.

\section{Action of the Dilatation Operator}

We want to evaluate the action of the dilatation operator\footnote{The dilatation operator for $\beta$ deformed ${\cal N}=4$ super Yang-MIlls theory
is derived in \cite{Roiban:2003dw}.}
\bea
  D_\gamma = -g_{YM}^2 {\rm Tr}(ZY\partial_Y\partial_Z + YZ\partial_Z\partial_Y
                       - e^{2\pi i\gamma} ZY\partial_Z\partial_Y - e^{-2\pi i\gamma} YZ\partial_Y\partial_Z)
\label{DilOp}
\eea
on the restricted Schur polynomial
\bea
  \chi_{R,(r,s)\alpha\beta}(Z,Y) &=& {1\over n! m!}\sum_{\sigma\in S_{n+m}}{\rm Tr}_{(r,s)\alpha\beta}(\Gamma^R(\sigma))
  Y^{i_1}_{i_{\sigma(1)}}\cdots Y^{i_m}_{i_{\sigma(m)}}Z^{i_{m+1}}_{i_{\sigma (m+1)}}\cdots Z^{i_{m+n}}_{i_{\sigma(m+n)}}\cr
  &\equiv& {1\over n! m!}\sum_{\sigma\in S_{n+m}}{\rm Tr}_{(r,s)\alpha\beta}(\Gamma^R(\sigma))
  {\rm Tr}(\sigma Y^{\otimes\, m}Z^{\otimes\, n})
  \label{restschurpoly}
\eea
$R\vdash m+n$ specifies an irreducible representation (irrep) of $S_{n+m}$, which $r\vdash n$ and $s\vdash m$ so that $(r,s)$
specifies an irrep of $S_n\times S_m$. 
In (\ref{restschurpoly}) ${\rm Tr}_{(r,s)\alpha\beta}$ is an instruction to trace only over a subspace of the full carrier
space of $R$; see \cite{dssi,Bhattacharyya:2008rb} for more details.
The evaluation of all four terms in (\ref{DilOp}) are similair and can be carried out as in \cite{DeComarmond:2010ie}.
We will explain in detail how to evaluate the first term
\bea
T_1 &=& {\rm Tr}(ZY\partial_Y\partial_Z)\chi_{R,(r,s)\alpha\beta}(Z,Y)\cr
    &=& Z^i_j Y^j_k {d\over dY^l_k}{d\over dZ^i_l} \chi_{R,(r,s)\alpha\beta}(Z,Y)\cr
    &=&{1\over (n-1)! (m-1)!}\sum_{\sigma\in S_{n+m}}{\rm Tr}_{(r,s)\alpha\beta}(\Gamma^R(\sigma))
          \delta^{i_1}_{i_{\sigma (m+1)}}(ZY)^{i_{m+1}}_{i_{\sigma (1)}}
          Y^{i_2}_{i_{\sigma(2)}}\cdots Y^{i_m}_{i_{\sigma(m)}}Z^{i_{m+2}}_{i_{\sigma (m+2)}}
          \cdots Z^{i_{m+n}}_{i_{\sigma(m+n)}}\cr
    &=& {1\over (n-1)! (m-1)!}\sum_{\psi \in S_{n+m}}\delta^{i_1}_{i_{\psi (1)}}{\rm Tr}_{(r,s)\alpha\beta}(\Gamma^R(\psi\, (1,m+1)\,))
  {\rm Tr}(\psi (1,m+1) Y^{\otimes\, m}Z^{\otimes\, n})\cr
\nonumber
\eea
The sum over $S_{n+m}$ can be reduced to a sum over $S_{n+m-1}$ by employing the reduction rule of \cite{rg,dssi}.
The result is
\bea
T_1 ={1\over (n-1)!(m-1)!}\sum_{\psi\in S_{n+m-1}\,\,\psi(1)=1}\sum_{R'}c_{RR'}{\rm Tr}_{(r,s)\alpha\beta}(\Gamma^{R'}(\psi)\Gamma^R(\,(1,m+1)\,))\cr
\times {\rm Tr}(\psi (1,m+1) Y^{\otimes\, m}Z^{\otimes\, n})
\eea
We would now like to express this as a linear combination of restricted Schur polynomials. This is most easily achieved by
using the identity
\bea
{\rm Tr}(\sigma Y^{\otimes\, m}Z^{\otimes\, n})=\sum_{T,(t,u)\gamma\delta}{d_T n! m!\over d_t d_u (n+m)!}
\chi_{T,(t,u)\gamma\delta}(\sigma^{-1})\chi_{T,(t,u)\delta\gamma}(Z,Y)
\eea
which has been proved in \cite{rajmike}. The result is
\bea
T_1 =\sum_{T,(t,u)\gamma\delta}{d_T n m\over d_t d_u (n+m)!}
\sum_{\psi\in S_{n+m-1}\,\,\psi(1)=1}\sum_{R'}c_{RR'}{\rm Tr}_{(r,s)\alpha\beta}(\Gamma^{R'}(\psi)\Gamma^R(\,(1,m+1)\,))\times\cr
\times\chi_{T,(t,u)\gamma\delta}(\, (1,m+1)\,\psi^{-1})\chi_{T,(t,u)\delta\gamma}(Z,Y)\cr
\eea
The fundamental othogonality relation can now be used to perform the sum over $\psi$. We finally obtain
\bea
T_1 = \sum_{T,(t,u)\gamma\delta}\sum_{R'} {c_{RR'} d_T n m\over d_t d_u d_{R'} (n+m)}\times\cr
{\rm Tr}_{R\oplus T} 
(\Gamma^R(\,(1,m+1)\,)P_{R,(r,s)\alpha\beta}I_{R'T'}P_{T,(t,u)\gamma\delta}\Gamma^T(\, (1,m+1)\,)I_{T'R'})
\times\chi_{T,(t,u)\delta\gamma}(Z,Y)
\eea
We have explicitely indicated that the trace appearing in this last expression is over the direct sum of
the carrier spaces of $R$ and of $T$.
All four terms in the dilatation operator can be treated in exactly the same way.

The relation between the restricted Schur polynomials given above and those with normalized two point function is
\bea
  \chi_{R,(r,s)\alpha\beta}(Z,Y)=\sqrt{f_R {\rm hooks}_R\over {\rm hooks}_r {\rm hooks}_s}O_{R,(r,s)\alpha\beta} (Z,Y)
\eea
Acting on normalized operators we have
\bea
 D_\gamma O_{R,(r,s)\alpha\beta}(Z,Y)=-g_{YM}^2\sum_{T,(t,u)\delta\gamma}\sum_{R'}
     {c_{RR'} d_T n m\over d_t d_u d_{R'} (n+m)}
\sqrt{f_T {\rm hooks}_T {\rm hooks}_r {\rm hooks}_s\over f_R {\rm hooks}_R {\rm hooks}_t {\rm hooks}_u}\times\cr
\times{\rm Tr}_{R\oplus T}\left[
\Gamma^R(\,(1,m+1)\,)P_{R,(r,s)\alpha\beta}I_{R'T'}P_{T,(t,u)\gamma\delta}\Gamma^T(\,(1,m+1)\,)I_{T'R'}\right.\cr
+P_{R,(r,s)\alpha\beta}\Gamma^R(\,(1,m+1)\,)I_{R'T'}\Gamma^T(\,(1,m+1)\,)P_{T,(t,u)\gamma\delta}I_{T'R'}\cr
-e^{-2\pi i\gamma}\Gamma^R(\,(1,m+1)\,)P_{R,(r,s)\alpha\beta}I_{R'T'}\Gamma^T(\,(1,m+1)\,)P_{T,(t,u)\gamma\delta}I_{T'R'}\cr
-e^{2\pi i\gamma}\left. P_{R,(r,s)\alpha\beta}\Gamma^R(\,(1,m+1)\,)I_{R'T'}P_{T,(t,u)\gamma\delta}\Gamma^T(\,(1,m+1)\,)I_{T'R'}\right]
O_{T,(t,u)\delta\gamma}\cr
\label{deformeddil}
\eea
The expression for the action of the one loop dilatation operator (\ref{deformeddil}) is exact to all orders in $1/N$.
The chief difficulty in evaluating (\ref{deformeddil}) explicitely is in constructing the operators $P_{R,(r,s)\alpha\beta}$ 
and in performing the trace over $R\oplus T$. Important recent progress on these issues has been achieved\cite{Koch:2011hb,mn}
by realizing that both of these tasks can be accomplished, at large $N$, by using the displaced corners approximations.
Denote the number of rows in the Young diagram labeling the restricted Schur polynomial by $p$.
To capture the large $N$ (but not planar!) limit we use the displaced corners approximation.
To understand what this approximation is and when it applies we will review relevant aspects of the construction of
$P_{R,(r,s)\alpha\beta}$. To subduce $r\vdash n$ from $R\vdash m+n$ we remove $m$ boxes from $R$.
Each box in the Young diagram $R$ can be assigned a factor which is equal to $N-i+j$ for the box in row 
$i$ and column $j$. In the displaced corners approximation the difference between the factors of any two 
boxes of the $m$ boxes removed, is of order $N$, whenever the removed boxes come from different rows.
In this limit the action of the symmetric group becomes particularly simple\cite{mn} and this is ultimately
the reason why it is useful to consider this limit. We now associate each removed box with a vector in a $p$ 
dimensional vector space $V_p$. Thus, the $m$ removed boxes associated with the $Y$'s define a vector in 
$V_p^{\otimes\, m}$. The trace over $R\oplus T$ now factorizes into a trace over $r\oplus t$ and a trace 
over $V_p^{\otimes m}$. The bulk of the work is in evaluating the trace over $V_p^{\otimes m}$. 
This trace can now be evaluated using the methods developed in \cite{Koch:2011hb}.
It is possible and useful to write the intertwining maps in terms of the basis for the 
fundamental representation of the Lie algebra u$(p)$ given by $(E_{ij})_{ab}=\delta_{ia}\delta_{jb}$ and obeying
\bea
  E_{ij}E_{kl} = \delta_{jk} E_{il}
\eea
A box is removed from row $i$ is associated to a vector $v_i$, which is an eigenstate of $E_{ii}$ with eigenvalue 1.
If we remove a box from row $i$ of $R$ and a box from row $j$ of $T$, assuming that $R'$ and $T'$ have the same shape, 
we have
\bea
  I_{T'R'}=E^{(1)}_{ji}
\eea
Denote the number of boxes in row $i$ of $R$ by $r_i$ and the number of boxes removed from
row $i$ of $R$ to obtain $r$ by $m_i$. We collect the $m_i$ into a vector $\vec{m}$ and say
that we remove $\vec{m}$ from $R$ to obtain $r$. We are now ready to evaluate the trace over 
$V_p^{\otimes m}$. 
Assume that we remove the box from row $i$ of $R$ to obtain $R'$, and from row $j$ of $T$ to obtain $T'$.
These terms have a coefficient of $\sqrt{c_{RR'}c_{TT'}}=\sqrt{(N+r_i)(N+r_j)}$.
The intertwining maps are $I_{T'R'}=E^{(1)}_{ji}$ and $I_{R'T'}=E^{(1)}_{ij}$.
We remove $\vec{m}$ from $R$ to obtain $r$ and $\vec{n}$ from $T$ to obtain $t$.
The four traces we need to evaluate are
\bea
{\rm Tr}\left(\Gamma^R(\,(1,m+1)\,)P_{R,(r,s)\alpha\beta}I_{R'T'}P_{T,(t,u)\gamma\delta}\Gamma^T(\,(1,m+1)\,)I_{T'R'}\right)\cr
={\rm Tr}\left(E^{(m+1)}_{ii}P_{R,(r,s)\alpha\beta}E^{(1)}_{ii}P_{T,(t,u)\gamma\delta}\right)\cr
=\sum_{k}\delta_{RT}\delta_{rt}\delta_{\vec{m}\vec{n}}d_{r'(i)}
\langle s,\beta,a |E^{(1)}_{ii}|u,\gamma,b\rangle\,\langle u,\delta,b|E^{(1)}_{kk}|s,\alpha,a\rangle
\eea
\bea
{\rm Tr}\left(P_{R,(r,s)\alpha\beta}\Gamma^R(\,(1,m+1)\,)I_{R'T'}\Gamma^T(\,(1,m+1)\,)P_{T,(t,u)\gamma\delta}I_{T'R'}\right)\cr
={\rm Tr}\left(E^{(1)}_{ii}P_{R,(r,s)\alpha\beta}E^{(m+1)}_{ii}P_{T,(t,u)\gamma\delta}\right)\cr
=\sum_{k}\delta_{RT}\delta_{rt}\delta_{\vec{m}\vec{n}}d_{r'(i)}
\langle s,\beta,a |E^{(1)}_{kk}|u,\gamma,b\rangle\,\langle u,\delta,b|E^{(1)}_{ii}|s,\alpha,a\rangle
\eea
\bea
{\rm Tr}\left(\Gamma^R(\,(1,m+1)\,)P_{R,(r,s)\alpha\beta}I_{R'T'}\Gamma^T(\,(1,m+1)\,)P_{T,(t,u)\gamma\delta}I_{T'R'}\right)\cr
=\sum_{kl}{\rm Tr}\left(E^{(1)}_{jk}E^{(m+1)}_{ki}P_{R,(r,s)\alpha\beta}E^{(1)}_{il}E^{(m+1)}_{lj}P_{T,(t,u)\gamma\delta}\right)\cr
=\delta_{r'(i)t'(j)}\delta_{\vec{m}\vec{n}} d_{r'(i)} 
\langle s,\beta,a |E^{(1)}_{ii}|u,\gamma,b\rangle\,\langle u,\delta,b|E^{(1)}_{jj}|s,\alpha,a\rangle
\eea
\bea
{\rm Tr}\left(P_{R,(r,s)\alpha\beta}\Gamma^R(\,(1,m+1)\,)I_{R'T'}P_{T,(t,u)\gamma\delta}\Gamma^T(\,(1,m+1)\,)I_{T'R'}\right)\cr
=\sum_{kl}{\rm Tr}\left(E^{(1)}_{li}E^{(m+1)}_{jl}P_{R,(r,s)\alpha\beta}E^{(1)}_{kj}E^{(m+1)}_{ik}P_{T,(t,u)\gamma\delta}\right)\cr
=\delta_{r'(i)t'(j)}\delta_{\vec{m}\vec{n}} d_{r'(i)} 
\langle s,\beta,a |E^{(1)}_{jj}|u,\gamma,b\rangle\,\langle u,\delta,b|E^{(1)}_{ii}|s,\alpha,a\rangle
\eea

Using these results we obtain the following action for the dilatation operator
\bea
D_\gamma O_{R,(r,s)\alpha\beta}=-g_{YM}^2\sum_{T,(t,u)\delta\gamma}[ M^{(1)}_{R,(r,s)\alpha\beta ;T,(t,u)\delta\gamma}
+M^{(2)}_{R,(r,s)\alpha\beta ;T,(t,u)\delta\gamma}]O_{T,(t,u)\delta\gamma}
\label{newdilaction}
\eea
where
\bea
M^{(1)}_{R,(r,s)\alpha\beta ;T,(t,u)\delta\gamma}&=&{m\over\sqrt{d_s d_u}}\delta_{\vec{m}\vec{n}}\sum_{j=2}^p\sum_{i=1}^{j-1}\Big[
\langle s,\beta,a |E^{(1)}_{ii}|u,\gamma,b\rangle\,\langle u,\delta,b|E^{(1)}_{jj}|s,\alpha,a\rangle\times\cr
&&[(2N+r_i+r_j)\delta_{RT}\delta_{rt}-\sqrt{(N+r_i)(N+r_j)}
(\delta_{T,R_{ij}^+}\delta_{t,r_{ij}^+} e^{2\pi i\gamma} + \delta_{T,R_{ij}^-}\delta_{t,r_{ij}^-} e^{-2\pi i\gamma})]\cr
%&&\times \langle s,\beta,a |E^{(1)}_{ii}|u,\gamma,b\rangle\,\langle u,\delta,b|E^{(1)}_{jj}|s,\alpha,a\rangle\cr
&&+[(2N+r_i+r_j)\delta_{RT}\delta_{rt}-\sqrt{(N+r_i)(N+r_j)}
(\delta_{T,R_{ij}^+}\delta_{t,r_{ij}^+} e^{-2\pi i\gamma} + \delta_{T,R_{ij}^-}\delta_{t,r_{ij}^-} e^{2\pi i\gamma})]\cr
&&\times \langle s,\beta,a |E^{(1)}_{jj}|u,\gamma,b\rangle\,\langle u,\delta,b|E^{(1)}_{ii}|s,\alpha,a\rangle
\Big]
\eea
and
\bea
M^{(2)}_{R,(r,s)\alpha\beta ;T,(t,u)\delta\gamma}={4m\sin^2 (\pi\gamma)\over\sqrt{d_s d_u}}\delta_{\vec{m}\vec{n}}
\delta_{RT}\delta_{rt}\sum_{i=1}^{p}(N+r_i)
\langle s,\beta,a |E^{(1)}_{ii}|u,\gamma,b\rangle\cr
\times\langle u,\delta,b|E^{(1)}_{ii}|s,\alpha,a\rangle
\eea
We have used $r_{ij}^+$ to denote the Young diagram obtained from $r$ by removing a box from $j$ and adding it to
row $i$, while $r_{ij}^-$ is obtained by removing a box from row $i$ and adding it to row $j$. Delta functions like
$\delta_{t,r_{ij}^+}$ are 1 if the two Young diagrams have the same shape and are zero otherwise. Notice that the action
of each of the two terms in the dilatation operator has factored into the product of two actions, one that
acts only on Young diagram $r$ (i.e. on the $Z$ fields) and another that acts only on the Young diagram $s$ (i.e. on the $Y$ fields).
Further, since the action of the second term on the $r$ label is trivial, we can diagonalize on the $s,\mu_1\mu_2; u\nu_1\nu_2$ and
the $R, r; T, t$ labels separately. This is identical to what happens in the undeformed case. In the undeformed case
it is the diagonalization on the $s,\mu_1\mu_2; u\nu_1\nu_2$ labels that is solved by the Gauss graph operators.

Our next task is to write the action of the dilatation operator in the Gauss graph basis.
The first term in (\ref{newdilaction}) has exactly the same form as the action of the dilatation operator 
in the undeformed theory, so that after using the results of \cite{DCI} we immediately obtain the action
of this term in the Gauss graph basis. We will thus focus on the second term in (\ref{newdilaction}). 
Introduce the notation
\bea
 D_\gamma^{(2)}O_{R,(r,s)\alpha\beta}=
-g_{YM}^2 {4m\delta_{\vec{m}\vec{n}}\sin^2 (\pi\gamma)\over \sqrt{d_s d_u}}\sum_{u\,\delta\,\gamma}\sum_i (N+r_i)
\langle s,\beta,a |E^{(1)}_{ii}|u,\gamma,b\rangle\,\cr
\times\langle u,\delta,b|E^{(1)}_{ii}|s,\alpha,a\rangle
O_{R,(r,u)\delta\gamma}
\nonumber
\eea
Recall that the Gauss graph operators are given by
\bea
  O_{R,r}(\sigma)={|H|\over\sqrt{m!}}\sum_{j,k}\sum_{s\vdash m}\sum_{\alpha\beta}\sqrt{d_s}
\Gamma^{s}_{jk}(\sigma)B^{s\to 1_H}_{j\alpha}B^{s\to 1_H}_{k\beta}O_{R,(r,s)\alpha\beta}
\eea
where $H$ is defined in terms of $\vec{m}$ by $H=S_{m_1}\times S_{m_2}\times\cdots\times S_{m_p}$,
$B^{s\to 1_H}_{j\alpha}$ are the branching coefficients from irrep $s$ of $S_m$ to the trivial of $H$
($j$ is a label for states in the carrier space of $s$ and $\alpha$ is a multiplicity label - distinguishing
the copies of the trivial irrep of $H$ subduced by $s$) and $\sigma$ is an element of the double coset
$H\setminus S_m /H$. We now compute
\bea
\langle O^\dagger_{T,t}(\sigma_2)D_\gamma^{(2)}O_{R,r}(\sigma_1)\rangle
=-\delta_{RT}\delta_{rt}4mg_{YM}^2\sin^2(\pi\gamma)\sum_i (N+r_i)\cr
\sum_{s,u\vdash m}\sum_{\alpha\beta\gamma\delta}\sum_i {|H|^2\over m!}
\langle s,\beta,a|E_{ii}^{(1)}|u,\gamma,b\rangle\langle u,\delta,b|E_{ii}^{(1)}|s,\alpha,a\rangle\cr
\Gamma^{(s)}_{jk}(\sigma_2)B^{s\to 1_H}_{j\alpha}B^{s\to 1_H}_{k\beta}
\Gamma^{(u)}_{lm}(\sigma_1)B^{u\to 1_H}_{l\delta}B^{u\to 1_H}_{m\gamma}\cr
\label{ingb}
\eea
Some algebra shows\cite{DCI}
\bea
\sum_u |u,\gamma,b\rangle\langle u,\delta,b| \Gamma^{(u)}_{lm}(\sigma_1)B^{u\to 1_H}_{l\delta}B^{u\to 1_H}_{m\gamma}
={1\over |H|^3}\sum_{\sigma,\tau\in S_m}\sum_{\gamma_1,\gamma_2\in H}
  \delta(\gamma_1\sigma_2^{-1}\gamma_2^{-1}\tau^{-1}\sigma)|v_\sigma\rangle\langle v_\tau|
 \nonumber
\eea
Consequently (\ref{ingb}) becomes
\bea
-{g_{YM}^2 |H|^2\over m!}{m\over |H^4|}\sum_{\beta ,\tau}\sum_{\gamma_2\, \gamma_4}
\langle\bar{v}|E_{ii}^{\beta^{-1}(1)}\beta^{-1}\tau\gamma_2\sigma_2|\bar{v}\rangle
\langle\bar{v}|E_{ii}^{\tau^{-1}(1)}\tau^{-1}\beta\gamma_4\sigma_1^{-1}|\bar{v}\rangle
\eea
Introduce the set $S_i$ of integers that lie in the range 
$m_1+m_2+\cdots m_{i-1}+1\le x\le m_1+m_2+\cdots m_{i-1}+m_i$.
Then (\ref{ingb}) becomes
\bea
-{g_{YM}^2 |H|^2\over m!}\sum_i (N+r_i){m\over |H|^4}\sum_{\beta\tau}\sum_{\gamma_i}
\delta(\beta^{-1}\tau\gamma_2 \sigma_2 \gamma_3)
\delta(\tau^{-1}\beta\gamma_4\sigma_1^{-1}\gamma_1)
\sum_{k,l\in S_i}\delta (\beta^{-1}(1),k)\delta (\tau^{-1}(1),l)
\nonumber
\eea
The last two delta functions in the above expression imply that $\beta(k)=\tau(l)$. Rewriting this expression
entirely in terms of $\beta^{-1}\tau$ we find
\bea
-g_{YM}^2 {1\over m!\,|H|^2}\sum_i(N+r_i)\sum_{\beta\,\,\tau}\sum_{\gamma_i}
\delta(\beta^{-1}\tau\gamma_2 \sigma_2 \gamma_3)
\delta(\tau^{-1}\beta\gamma_4\sigma_1^{-1}\gamma_1)
\sum_{k,l\in S_i}\delta (\beta^{-1}\tau(k),l)\cr
=-g_{YM}^2 {1\over |H|^2}\sum_i(N+r_i)\sum_{\beta}\sum_{\gamma_i}
\delta(\beta^{-1}\gamma_2 \sigma_2 \gamma_3)
\delta(\beta\gamma_4\sigma_1^{-1}\gamma_1)
\sum_{k,l\in S_i}\delta (\beta^{-1}(k),l)\cr
=-g_{YM}^2 {1\over |H|^2}\sum_i(N+r_i)\sum_{\beta}\sum_{\gamma_i}
\delta(\beta^{-1}\gamma_2 \sigma_2 \gamma_3)
\delta(\beta\gamma_4\sigma_1^{-1}\gamma_1)
n_{ii} (\beta^{-1})\cr
=-g_{YM}^2 \sum_i(N+r_i){1\over |H|^2}\sum_{\gamma_i}
\delta(\gamma_2 \sigma_2 \gamma_3\gamma_4\sigma_1^{-1}\gamma_1)
n_{ii} (\gamma_4\sigma_1^{-1}\gamma_1)\cr
=-g_{YM}^2\sum_i(N+r_i)\sum_{\gamma_1\, \gamma_2}\delta (\gamma_1\sigma_2\gamma_2\sigma_1^{-1})n_{ii}(\sigma_1)
\cr
\eea
Notice that $\sum_{\gamma_1\, \gamma_2}\delta (\gamma_1\sigma_2\gamma_2\sigma_1^{-1})$ is the delta
function on the double coset. Thus, the new term is diagonal in the Gauss graph basis. 
This term is an additive constant (since it is diagonal in the $R,r$ labels and the elements on the 
diagomal do not depend on $r$ or $R$)
\bea
\langle O^\dagger_{T,t}(\sigma_2)D_\gamma^{(2)}O_{R,r}(\sigma_1)\rangle
=-g_{YM}^2\delta_{RT}\delta_{rt}\sum_i(N+r_i)n_{ii}(\sigma_1)\sum_{\gamma_1\, \gamma_2}\delta (\gamma_1\sigma_2\gamma_2\sigma_1^{-1})
\eea

Next, introduce the notation
\bea
 D_\gamma^{(1)}O_{R,(r,s)\alpha\beta}
=-g_{YM}^2\sum_{T,(t,u)\delta\gamma} M^{(1)}_{R,(r,s)\alpha\beta ;T,(t,u)\delta\gamma}O_{T,(t,u)\delta\gamma}\cr
\eea
Using the results from \cite{DCI}, we find
\bea
&&\langle O^\dagger_{T,t}(\sigma_2)D_\gamma^{(1)}O_{R,r}(\sigma_1)\rangle =
-g_{YM}^2\sum_{\gamma_1 ,\gamma_2\in H}\delta (\gamma_1\sigma_2\gamma_2\sigma_1^{-1})\sum_{j=2}^p\sum_{i=1}^{j-1}\cr
&&\quad\Big\{ n_{ij}^-(\sigma_1)[(2N+r_i+r_j)\delta_{RT}\delta_{rt}-\sqrt{(N+r_i)(N+r_j)}
(\delta_{T,R_{ij}^+}\delta_{t,r_{ij}^+} e^{2\pi i\gamma} + \delta_{T,R_{ij}^-}\delta_{t,r_{ij}^-} e^{-2\pi i\gamma})]\cr
&&\quad +n_{ij}^+ (\sigma_1) [(2N+r_i+r_j)\delta_{RT}\delta_{rt}-\sqrt{(N+r_i)(N+r_j)}
(\delta_{T,R_{ij}^+}\delta_{t,r_{ij}^+} e^{-2\pi i\gamma} + \delta_{T,R_{ij}^-}\delta_{t,r_{ij}^-} e^{2\pi i\gamma})]\Big\}\cr
\eea
In the above expression $n_{ij}^+$ is the number of strings going from $i$ to $j$; in terms of the double coset
element we have
\bea
   n_{ij}^+(\sigma)=\sum_{k\in S_i}\sum_{l\in S_j}\delta (\sigma(k),l)
\eea
$n_{ij}^-$ is the number of strings going from $j$ to $i$ and in terms of the double coset element 
\bea
   n_{ij}^-(\sigma)=\sum_{k\in S_i}\sum_{l\in S_j}\delta (\sigma(l),k)
\eea
This completes the evaluation of the action of the deformed dilatation operator in the Gauss graph basis.

\section{Continuum Limit of $D_\gamma$}

To obtain the spectrum of anomalous dimensions we still have to solve the eigenproblem on the $R, r; T, t$ labels
which amounts to solving a difference equation.
At large $N$ we can take a continuum limit that replaces the diference equation with a partial differential equation. 
In the undeformed case this partial differential equation describes a system of particles 
interacting with a quadratic pairwise interaction potential. 
In this section we will consider the continuum limit of the $R, r; T, t$ eigenproblem in the deformed case.

In the large $N$ limit, for the operators we consider, we know that $N+r_0\to\infty$. Set $r_i=r_0 + l_i$ for $i>0$. 
To obtain the continuum limit we will fix the variables
$x_i={l_i\over\sqrt{N+r_0}}$
to be order 1. To keep track of the variables $l_i$ in what follows, we introduce a new notation for the operators $O_{R,r}(\sigma)$ denoting
them $O(\sigma,r_0,l_1,...,l_{p-1})$.
Assume that the operators of a good scaling dimension have the form
\bea
  O=\sum_{r_0,l_1,\cdots,l_{p-1}}f(r_0,l_1,...,l_{p-1})O(\sigma,r_0,l_1,...,l_{p-1})
\eea
The eigenvalue problem $DO=\Gamma O$ implies that
\bea
g_{YM}^2\sum_{i=1}^p\sum_{j=i+1}^p n_{ij}^-[(2N+r_i+r_j)f(r_0,l_1,...,l_{p-1})\cr
-\sqrt{(N+r_i)(N+r_j)}(e^{2\pi i\gamma}f(...,l_i+1,...,l_j-1,...) +e^{-2\pi i\gamma}f(...,l_i-1,...,l_j+1,...))]\cr
+g_{YM}^2\sum_{i=1}^p\sum_{j=i+1}^p n_{ij}^+[(2N+r_i+r_j)f(r_0,l_1,...,l_{p-1})\cr
-\sqrt{(N+r_i)(N+r_j)}(e^{-2\pi i\gamma}f(...,l_i+1,...,l_j-1,...) +e^{+2\pi i\gamma}f(...,l_i-1,...,l_j+1,...))]\cr
+4g_{YM}^2\sin^2 (\pi\gamma) \sum_{i=1}^p (N+r_i)n_{ii}f(r_0,l_1,...,l_{p-1}) =\Gamma f(r_0,l_1,...,l_{p-1})
\eea
Now, make use of the expansions
\bea
  \sqrt{(N+r_0+l_i)(N+b_0+l_j)} = N+r_0 +{x_i+x_j\over 2}\sqrt{ N + r_0}-{(x_i-x_j)^2\over 8}+...
\eea
and
\bea
  f(r_0,...,l_i-1,...,l_j+1,...)\to f(r_0,...,x_i-{1\over\sqrt{N+b_0}},...,x_j+{1\over\sqrt{N+b_0}},...)\cr
  =f(r_0,...,l_i,...,l_j,...)-{1\over\sqrt{N+r_0}}{\partial f\over\partial x_i}+{1\over\sqrt{N+b_0}}{\partial f\over\partial x_j}
+{1\over 2(N+b_0)}{\partial^2 f\over\partial x_i^2}\cr
+{1\over 2(N+ r_0)}{\partial^2 f\over\partial x_j^2}
-{1\over N+ r_0}{\partial^2 f\over\partial x_i\partial x_j}+...
\eea
to find
\bea
g_{YM}^2\sum_{i=1}^p\sum_{j=i+1}^p n_{ij}^-\Big[(2N+r_i+r_j)(1-\cos (2\pi\gamma))f(r_0,l_1,...,l_{p-1})\cr
-i\sin (2\pi\gamma)(2\sqrt{N+r_0}+x_i+x_j)\left({\partial f\over\partial x_i}-{\partial f\over\partial x_j}\right)\cr
+\cos(2\pi \gamma)\left( {(x_i-x_j)^2\over 4}-\left({\partial\over\partial x_i}-{\partial\over\partial x_j}\right)^2\right)f\Big]\cr
+g_{YM}^2\sum_{i=1}^p\sum_{j=i+1}^p n_{ij}^+\Big[(2N+r_i+r_j)(1-\cos (2\pi\gamma))f(r_0,l_1,...,l_{p-1})\cr
+i\sin (2\pi\gamma)(2\sqrt{N+r_0}+x_i+x_j)\left({\partial f\over\partial x_i}-{\partial f\over\partial x_j}\right)\cr
+\cos(2\pi \gamma)\left( {(x_i-x_j)^2\over 4}-\left({\partial\over\partial x_i}-{\partial\over\partial x_j}\right)^2\right)f\Big]\cr\cr
+4g_{YM}^2\sin^2 (\pi\gamma) \sum_{i=1}^p (N+r_i)n_{ii}f(r_0,l_1,...,l_{p-1}) =\Gamma f(r_0,l_1,...,l_{p-1})\cr
\label{togetspectrum}
\eea
This gives the partial differential equation that must be solved to obtained anomalous dimensions for the deformed theory. 
A comment is in order. 
In the undeformed case, configurations with $n_{ij}= 0$ and $n_{ii}\ne 0$ correspond to BPS operators. 
One of the implications of this is that any excitation of a single giant graviton (i.e. any restricted Schur polynomial built using only
$Z$s and $Y$s, labeled by Young diagrams that have only a single row or column) are BPS.
In the deformed case we see that this is clearly not so - $n_{ii}\ne 0$ leads to operators that are not BPS. 
Consequently, in the deformed case the excitations of a single giant graviton are not BPS.

\section{Spectrum}

The goal of this section is to compute the spectrum of the deformed dilatation operator by solving the eigenproblem (\ref{togetspectrum}).
Introduce a new set of coordinates $y_i=\sqrt{N+r_0}+x_i$.
In terms of these coordinates, rewrite (\ref{togetspectrum}) as
\bea
  Hf =\tilde{\Gamma}f
\eea
where
\bea
  H= g_{YM}^2 \sum_{i=1}^p\sum_{j=i+1}^p\left[ n_{ij}\cos(2\pi\gamma)(P_{ij}^2+{1\over 4}y_{ij}^2)
     +(n_{ij}^+-n_{ij}^-)\sin (2\pi\gamma)\,(y_i+y_j)\, P_{ij} \right]
\eea
\bea
  P_{ij}=i{\partial\over\partial y^i}-i{\partial\over\partial y^j}\qquad y_{ij}=y_i-y_j
\eea
\bea
  \tilde{\Gamma}= -2g_{YM}^2\sum_{i=1}^p\sum_{j=i}^p n_{ij}(N+r_i+N+r_j)\sin^2 (\pi\gamma)+\Gamma
\eea
Notice that $y_i+y_j$ commutes with $H$ and hence is a constant of the motion. It thus makes sense to 
shift $P_{ij}\to P_{ij}+\alpha (y_i+y_j)$. Indeed, by choosing
\bea
 \alpha ={(n_{ij}^--n_{ij}^+)\tan (2\pi\gamma)(y_i+y_j)\over 2n_{ij}}
\eea
we find
\bea
  H= g_{YM}^2 \sum_{i=1}^p\sum_{j=i+1}^p\left[ n_{ij}\cos(2\pi\gamma)(P_{ij}^2+{1\over 4}y_{ij}^2)
     -{(n_{ij}^+-n_{ij}^-)^2\sin^2 (2\pi\gamma)\,(y_i+y_j)^2\over  4n_{ij}\cos (2\pi\gamma)} \right]
\eea
The second term inside the square braces commutes with the Hamiltonian and is thus a constant. Noting that
\bea
  \big[ P_{ij},{y_i-y_j\over 2}\big] = i
\eea
it is clear that $H$ is equivalent to a (shifted) Harmonic oscillator. The spectrum of the oscillator clearly depends 
on the deformation parameter $\gamma$.

\section{Discussion}

In this article we have computed the spectrum of anomalous dimensions in the Leigh-Strassler deformed ${\cal N}=4$ super Yang-Mills theory. 
The operators that we have studied are AdS/CFT dual to systems of giant gravitons.
This implies that although we work at large $N$, we are not in the planar limit of the theory.

A key motivation for this work has been to test the validity of the double coset ansatz within the deformed theory. 
We have found that the action of the dilatation operator continues to factorize into an action on the impurity labels 
$s\mu_1\mu_2;u\nu_1\nu_2$ associated to the $Y$ fields and an action on the labels $R,r;T,t$ associated to the $Z$ fields.
The deformed dilatation operator picks up an extra term as compared to the undeformed case. 
The extra term is diagonal in the Gauss graph basis so that the double coset ansatz continues to diagonalize the impurity labels.
This matches our expectations, motivated by the observation that the Gauss Law, which is closely tied to the double coset ansatz,
is a general feature expected of the AdS/CFT dual theory of the open string excitations of the giant graviton systems we consider.

We have also considered the diagonalization problem on the $R,r;T,t$ labels associated to the $Z$ fields. 
It turns out that this problem can again be reduced to a set of decoupled oscillators.
The deformed dilatation operator picks up an additional term as compared to the undeformed case.
This extra term produces an extra shift in the amomalous dimension.
The shift is positive as it should be. Indeed, a negative shift would produce operators with a dimension less than their
${\cal R}$-charge which is not possible in a unitary conformal field theory.
This predicts that all excitations of the giant gravitons in the deformed theory are not BPS.
As an example, for a system of $p=2$ giant gravitons we have ($k$ is any non-negative integer)
\bea
  \Gamma_k = 4g_{YM}^2 (N+r_1)n_{11}\sin^2 (\pi\gamma)+4g_{YM}^2 (N+r_2)n_{22}\sin^2 (\pi\gamma)\cr
             +2g_{YM}^2 (2N+ n)n_{12}\sin^2 (\pi\gamma)
             +4g_{YM}^2 n_{12}\cos (2 \pi\gamma)k
\eea
Setting $\gamma=0$ we recover the anomalous dimensions of the undeformed theory \cite{Carlson:2011hy}.

We have not considered comparing with dual AdS/CFT predictions\cite{Lunin:2005jy,Pirrone:2006iq,Hamilton:2006ri}.  
As just commented, since our operators are not BPS their anomalous dimensions are not protected quantities. 
Since the AdS/CFT duality is a strong/weak coupling duality, a direct comparison is almost sure to fail.
More precisely, the dual gravitational system is defined in the limit of large t' Hooft coupling $\lambda$ and small
$\gamma$ ($\gamma^2 \lambda$ is fixed) while our field theory computation is valid when $\lambda$ is small and
$\gamma$ is arbitrary.
However, since the quantum numbers of our operators become parametrically large with $N$, a comparison may still be 
possible\cite{Berenstein:2002jq,Gubser:2002tv,Frolov:2002av,Frolov:2003qc,Frolov:2003tu}. 
We leave this interesting question for the future.

{\vskip 1.0cm}

\noindent
{\it Acknowledgements:} RdMK and NN are supported by the South African Research Chairs
Initiative of the Department of Science and Technology and the National Research Foundation.
JM is supported by the National Research Foundation under the Thuthuka and Incentive Funding 
for Rated Researchers Program.
Any opinion, findings and conclusions or recommendations expressed in this material
are those of the authors and therefore the NRF and DST do not accept any liability
with regard thereto.

\end{document}